\documentstyle[multicol,aps,prb,epsf]{revtex}

\def\lsim{\lower -0.3ex \hbox{$<$} \kern -0.75em \lower 0.7ex \hbox{$\sim$}}
\def\gsim{\lower -0.3ex \hbox{$>$} \kern -0.75em \lower 0.7ex\hbox{$\sim$}}
\newcommand{\Vec}[1]{\mbox{\boldmath$#1$}}

\begin{document}

\draft

\title{Phase Diagram for the Hofstadter butterfly and 
integer quantum Hall effect in three dimensions}
\author{M. Koshino, H. Aoki}
\address{Department of Physics, University of Tokyo, Hongo, Tokyo
113-0033, Japan}
\author{T. Osada}
\address{Institute for Solid State Physics, University of Tokyo,
Kashiwa, Chiba 277-8581, Japan}
\author{K. Kuroki}
\address{Department of Applied Physics and
Chemistry, University of Electro-Communications, Chofu, Tokyo 182-8585, Japan}
\author{S. Kagoshima}
\address{Department of Basic Science, University of Tokyo, Komaba, Tokyo 
153-8902, Japan}

\date{\today}

\maketitle

\begin{abstract}
We give a perspective on the Hofstadter butterfly (fractal energy 
spectrum in magnetic fields), which we have shown to arise specifically in 
three-dimensional(3D) 
systems in our previous work.  (i) We first 
obtain the `phase diagram' on a parameter space of the transfer 
energies and the magnetic field for the appearance of Hofstadter's 
butterfly spectrum in anisotropic crystals in 3D.  
(ii) We show that the orientation of the external magnetic field can be 
arbitrary to have the 3D butterfly.  
(iii) We show that the butterfly is  beyond the semiclassical description.  
(iv) The required magnetic field for a representative organic metal 
is estimated to be modest ($\sim 40$ T) if we adopt higher Landau levels 
for the butterfly.  
(v) We give a simpler way of deriving the topological invariants that represent the quantum Hall numbers (i.e., two Hall conductivity 
in 3D, $\sigma_{xy}, \sigma_{zx}$, in units of $e^2/h$).
\end{abstract}


\begin{multicols}{2}
\narrowtext

\section{Introduction}

The problem of ``commensurability" in almost periodic
systems has been a source of fascination in various branches of physics.
In this respect, the Bloch states in periodic systems in strong
magnetic fields have especially appealing.  Namely, the electron system
with a periodic potential (or a tight-binding electron
system on a lattice) in two spatial dimensions(2D) exhibits,
when a uniform magnetic field is applied,
a self-similar energy spectrum, known as Hofstadter's
butterfly, with fractal wave functions\cite{Hofs}.
In this case, the problem is how the Bloch electrons
react to an external periodicity introduced by the
magnetic field.
This problem, raised more than twenty years ago by Hofstadter,
has been intensively investigated.

The special assets of this problem are:
(i) Mathematically, Schr\"{o}dinger's equation reduces to
Harper's equation with doubly-periodic potential, where
one periodicity is externally governed by the
strength of the applied magnetic field, $\Vec{B}$, so that
the Bloch states are controlled {\it in situ} by varying $B$.

(ii) In Hofstadter's butterfly (the fractal energy spectrum versus
$B$), each Landau band splits into fractal
subbands, where the integer quantum Hall effect(IQHE)
is known to occur.   The general argument
of Laughlin \cite{Laug} tells us that each of the fractally
split bands should carry
a quantized Hall current, while in usual cases
each Landau band does. Thouless {\it et al}\cite{Thou}
in fact have shown, by actually deriving the general formula for
the quantized Hall conductance in 2D periodic system without any special
assumption on the periodic potential, that
the quantum number characterizing the Hall current
carried by each subband is a topological invariant.

Now, for both of (i) and (ii) above, two dimensionality is
essential, since, first, Schr\"{o}dinger's equation reduces to
doubly-periodic equation only in 2D (while we have a triply-periodic
potential in 3D) and, second, the quantum Hall effect is usually considered
to be specific to 2D.  Nevertheless, we want to ask ourselves here: can
we have similar phenomena that survive, or even arise specifically, in
three dimensions (3D)?  In the last decade several authors have
indeed attempted to extend Hofstadter's problem to 3D
systems\cite{Hase,Kuns}.  Usually, however, subbands overlap in
energy or touch with each other in 3D, and fractal energy spectra have not
been identified. On the other hand, Halperin\cite{Halp} has shown,
following the line of the analysis for 2D, that if the spectrum has an
energy gap, IQHE should occur even in 3D.  Montambaux and
Kohmoto\cite{Mont} then proposed that 3D Bloch system in magnetic fields
can realize this situation to derive the formula for Hall
conductivities.  These results are generalized by Kohmoto, Halperin and
Wu\cite{Kohm}.

In our previous letter\cite{Kosh}, we have shown
that an analogue of Hofstadter's
butterfly does indeed arise in an anisotropic tight-binding lattice
$(t_x \gg t_y, t_z)$ when a
magnetic field is tilted, where
the conductive axis is denoted as $x$.  We have calculated the Hall
conductivity and have shown that each of $\sigma_{xy}, \sigma_{zx}$,
identified to be topological invariants,
exhibits an integer quantum Hall effect, with a curious
one-to-one correspondence with the QHE in 2D butterfly.
This has been explained with a
mapping between 2D and 3D.  A crucial point is
that the `3D butterfly' is specific to
three dimensionality rather than a remnant of the 2D butterfly.

The purpose of the present paper is to address the following new points: \par
\noindent (a)
to obtain a `phase diagram' that indicates
which region in $(t_y, t_z)$ space should exhibit
the butterfly.

\noindent (b)
to show that the 3D butterfly is general enough to
allow arbitrary orientation of the magnetic field.

\noindent (c) to clarify the relation with the semiclassical
quantization picture in magnetic fields in discussing the details of the
2D-3D mapping.

So the paper is organized as follows.
In Sec.\ref{SecPhase}, we review Harper's equations for the 3D
Bloch electrons in magnetic fields, and introduce the phase diagram of
3D butterfly.  In Sec.\ref{SecBx}, we extend the 2D-3D mapping to the
arbitrary orientation of the magnetic field to discuss the influence of
$B_x \ne 0$ on the butterfly spectrum.  In Sec.\ref{SecSemiCl}, we
compare our results with the semiclassical approximation.  In Sec.VI, we
turn to QHE in the 3D butterfly, where the Hall conductivity is calculated
in a much simpler way than in our earlier paper.  Experimental
possibilities to realize the 3D butterfly
is discussed in Sec.\ref{SecHall}, and the conclusion is given in
Sec.\ref{SecConcl}.

\section{The Condition for 3D butterfly}
\label{SecPhase}

\subsection{The 2D-3D mapping}

Let us first summarize the results obtained in our previous work,\cite{Kosh} 
where we first suggested that we can have a butterfly spectrum in 3D systems. 
We take a noninteracting tight-binding
electron system in a uniform magnetic field, $\Vec{B}$.
Schr\"{o}dinger's equation is
\begin{equation}
-\sum_j t_{ij} e^{i\theta_{ij}}\psi_j = E\psi_i \label{Harp}.
\end{equation}
Here $\psi_i$ is the wave function at site $i$,
the summation is over nearest-neighbor sites, 
and 
$\theta_{ij} = \frac{e}{\hbar}\int_{j}^{i}\Vec{A}\cdot {\rm d}\Vec{l}$ 
the Peierls phase factor arising from the magnetic flux 
where \Vec{A} is the vector potential with $\nabla\times\Vec{A} = \Vec{B}$. 

We consider a 3D simple-cubic lattice in a magnetic field
$\Vec{B}=(0,B\sin\theta ,B\cos\theta)$ assumed to be parallel
to the $yz$ plane.
If we choose the vector potential as
$\Vec{A}=(0,Bx\cos\theta,-Bx\sin\theta)$, 
we can substitute  $\psi$ in eq.(\ref{Harp}) with
$\psi_{lmn} = e^{i \nu_y m + i \nu_z n} F_l$, 
where $l,m,n$ are the site indexes along $x,y,z$, respectively.
Then we have
\begin{eqnarray}
  -t_x(F_{l-1} \,+\, F_{l+1}) \,-\,  [2t_y \cos(2\pi \phi_z l
  +\nu_y) \nonumber \\
  + 2t_z \cos(-2\pi \phi_y l +\nu_z)] F_l = E F_l
  \label{Harp3D}.
\end{eqnarray}
Here $t_x, t_y, t_z$ are the transfer integrals between nearest
neighbors along $x,y,z$, respectively, and $\phi_y$ or $\phi_z$ are
the numbers of flux quanta penetrating the facet of the unit cell normal
to $y$ or $z$ (Fig.\ref{geom}).  
Let us call two cosine potentials in
(\ref{Harp3D}) $V^{(1)}$ and $V^{(2)}$.

If we assume $\phi_y, \phi_z \ll 1$, the periods of 
these cosine potentials are much greater 
than the original lattice constant, so that the problem reduces to 
a continuous 1D system if we apply the
effective-mass approach.  One can then regard the potential minima of
$V^{(1)}$ as effective `sites', which we call the wells to distinguish
from the original sites.  Each well feels the residual potential
$V^{(2)}$.  If the well is deep enough, several bound
states appear at each well, where each state forms a tight-binding band,
which corresponds to a Landau band.  

Let us define $t'$ as the hopping
between adjacent wells, where $4t'$ gives 
the width (Harper's broadening) of a
Landau level.  Then eq.(\ref{Harp3D}) reads, in terms of the
wells,
\begin{eqnarray}
  -t'(J_{l'-1} \,+\, J_{l'+1}) \,-\, 2t_z \cos[-2\pi (\phi_y /
  \phi_z)l' \nonumber \\ + (\phi_y / \phi_z)\nu_{y} + \nu_z] J_{l'} =
  E J_{l'}\label{HarpEff},
\end{eqnarray}
where $J_{l'}$ is the amplitude at the well $l'$.
We can see that the equation
has the same form as that of the 2D system 
(eq.(\ref{Harp3D}) with $t_z =0$), which explains why 
a butterfly spectrum can appear also in 3D.
If we recall that a 2D system exhibits a butterfly 
when isotropic ($t_x \approx t_y$), 
one can predict that the Landau band for which 
\begin{equation}
t' \approx t_z
\label{tprimetz}
\end{equation}
should exhibit a butterfly structure, where the 
value of $t'$ varies from one Landau band to another.


\subsection{The `phase diagram' for the 3D butterfly}

We have to satisfy some conditions on 
$(t_x,t_y,t_z,\phi_y,\phi_z)$ to have the 3D butterfly.
By systematically examining them we can obtain the `phase diagram',
which was not obtained in our previous letter.  We again assume
$\phi_z,\phi_y \ll 1$, and
we take $t_x=1$ as the unit of energy.

Let us begin with the pure 2D case ($t_z = 0$), where we have only
one potential term $V^{(1)}$ in Harper's equation.  If one
approximates a well in $V^{(1)}$ as a harmonic potential, the energy is
quantized into $\hbar\omega = 4\pi \phi_z \sqrt{t_y}$.  The
condition that there exist more than one levels bound to each
well (i.e. Landau levels) is then $\frac{1}{2}\hbar\omega < 4t_y$ (depth
of the well), i.e.,
\begin{equation}
\pi\phi_z < 2\sqrt{t_y}.
\label{cond1}
\end{equation}
If the opposite inequality holds, 
$V^{(1)}$ has only a weak effect on the
energy spectrum and wave functions. 

If we turn $t_z$ on, each Landau level feels the residual potential
$V^{(2)}$ and can split into butterfly bands.
For the butterfly gaps to be resolved, however, 
$V^{(2)}$ must be weak enough so that
it does not mix different Landau levels.  In other words, $4t_z$,
the amplitude of $V^{(2)}$, must be smaller than $\hbar \omega = 4\pi
\phi_z \sqrt{t_y}$, or
\begin{equation}
t_z < \pi \phi_z \sqrt{t_y}.
\label{cond2}
\end{equation}

We can now construct the phase diagram.  
Since an equivalent discussion can be made when 
we interchange $y$ and $z$, we must add the conditions
\begin{eqnarray}
\pi\phi_y < 2\sqrt{t_z} \label{cond3}\\
t_y < \pi \phi_y \sqrt{t_z}\label{cond4}.
\end{eqnarray}
The resultant phase diagram is shown in Fig.\ref{phase}.  
The phase diagram consists 
of 1D, 2D, 3D butterfly, and 3D regions.  
``1D" is the region where neither
eq.(\ref{cond1}) nor eq.(\ref{cond3}) are satisfied.  
There, both $V^{(1)}$ and $V^{(2)}$ are too weak and the energy spectrum
can only have Bragg-reflection gaps.  
Since this phase is dominated by $t_x$,
we call this 1D-like.  An example, with a Bragg gap, 
is depicted in Fig.\ref{1Dlike}.  
``2D-like" is the region where
only one of eq.(\ref{cond1}) or eq.(\ref{cond3}) is satisfied. There, one of
$t_y$ or $t_z$ has only minor effect and the spectrum is
2D-like in that we have only (Harper-broadened) Landau levels.
Then we come to the ``3D butterfly" phases, where 
eqs.(\ref{cond1}),(\ref{cond2}) (or eqs.(\ref{cond3}),(\ref{cond4}) 
are satisfied.  
The remaining region is ``3D", where all the three hoppings are too
large for energy gaps to survive.  Namely, 
even when both eq.(\ref{cond1}) and eq.(\ref{cond3}) are satisfied, the
butterfly gaps are wiped out when the other conditions are not
satisfied, i.e., when the residual potential is too large. 
So the 3D butterfly is allowed to appear somewhere 
in between the 2D and 3D phases.
When the magnetic field is too small, two parabolic phase 
boundaries coalesce to 
the axes and the 3D phase becomes dominant.

We display typical energy spectrum in each phase in Fig.\ref{btflMany}.
In this figure, panels (b)(c)(d) all belong to the 3D-butterfly 
phase, but we can immediately notice that the Landau band on 
which the butterfly appears varies according to where we sit 
on the phase diagram.  This is because within this phase 
the 3D butterfly appears when $t' \approx t_z$ (eq.\ref{tprimetz}), 
namely, only the 
Landau level (not necessarily the lowest) that has 
the right width exhibits the butterfly.  


\subsection{Estimate of the effective transfer integral}
\label{SecEstim}
 
In our previous letter, we have numerically estimated 
the Landau-level broadening $t'$ for the lowest
Landau band to predict when 3D butterflies appear.
Here we give a derivation of the scaling law 
for $t'$ introduced there.

Consider first an electron in one dimension in the potential $V^{(1)} =
-2t_y \cos(2\pi \phi_{z}l)$ (no $V^{(2)}$) in eq.(\ref{Harp3D}).  Here we
have omitted the phase factor $\nu_y$, which has no effect on $t'$.  Our
aim is to express $t'$ as a function of $(t_x,t_y,\phi_z)$.  Harper's
equation in this case is
\begin{equation}
-t_x(F_{l-1} \,+\, F_{l+1}) \,-\, 2t_y \cos(2\pi \phi_{z}l) F_{l} = E F_{l}.
\end{equation}
We have assumed the period of $V^{(1)}$ is much greater than the lattice
constant, so the equation can be cast into a differential
equation up to a constant,
\begin{equation}
-\frac{{\rm d^2} F}{{\rm d}l^2} \,-\, 2t_y \cos(2\pi \phi_{z}l) F(l) = E F(l).
\end{equation}
We again take $t_x=1$ as the  unit of energy.  
If we measure the length in the well spacing by introducing 
\begin{equation}
\tilde{l} \equiv 2\pi \phi_{z}l,
\end{equation}
the equation becomes
\begin{equation}
-\frac{{\rm d^2} F}{{\rm d}\tilde{l}^2} \,-\, 2\frac{t_y}{(2\pi\phi_z)^2} \cos(\tilde{l}) F(\tilde{l}) = \tilde{E}F(\tilde{l}),
\end{equation} 
where $\tilde{E}= E/(2\pi\phi_z)^2$ is the rescaled energy.
This equation indicates that the form of the wave
function depends only on $t_y / \phi_z^2$.

Since $t'$ depends on the overlap of the neighboring 
wave functions and the amplitude of the periodic potential $V^{(1)}$, 
$t'$ can be written as
\begin{equation}
t' = t_y \, \, f \left( \frac{t_y}{\phi_z^2} \right),
\label{EffectiveT}
\end{equation}
where the function $f$, which only depends on the 
Landau index, represents the overlap of 
the neighboring wave functions.  
Hence we have only to calculate $t'(t_y)$ for one value of $\phi_z$.  
In the limit of large barrier between neighboring
wells, $(t_y / \phi_z^2) \rightarrow \infty$, $f$ vanishes 
exponentially. In the opposite limit 
of $(t_y / \phi_z^2)\rightarrow 0$, wells
become shallow, so that the transfer to the second- (and more distant)
neighbor transfers cannot be neglected, and the approximation should fail.

This argument can be used for each of the Landau levels.
Higher levels have larger $t'$ since the wave function bound in
the well is more extended.  
So, taking account of the clear butterfly condition $t' \approx t_z$,
we find that as $t_z$ is increased the Landau level in which 
butterfly appears becomes higher, as we have seen in Fig.\ref{btflMany}.

\section{3D butterfly in an arbitrary orientation of $\Vec{B}$}
\label{SecBx}

So far we have assumed that the magnetic field lies in the plane ($yz$) 
normal to 
the conductive axis.  So the next question is whether we can relax this
condition for realizing the butterfly.  About ten years ago, several
authors investigated 3D Hofstadter problem in arbitrarily oriented
magnetic fields\cite{Hase,Kuns}.  They have shown that, even if
each of three components of magnetic field is not zero,
Schr\"{o}dinger's equation can be transformed into a one dimensional
tight-binding problem with distant-neighbor hoppings.  Namely, if we
choose a vector potential that depends on both $y$ and $z$, the
equation, reduced to a 1D tight-binding equation,
has $n_y$- and $n_z$-distant hoppings where $n_z/n_y = \phi_z/\phi_y$ with
$n_y,n_z$ being coprime integers. We can adopt this gauge to
calculate the 3D butterfly for $B_x \ne 0$ here.

The result for the energy spectra when the magnetic field is tilted away 
from the $yz$ plane ($B_x \neq 0$)
is shown in Fig.\ref{btflBx}.
Surprisingly, the spectrum is almost identical to that 
for $B_x = 0$ (Fig.\ref{btflBx}(a)).
In other words, the magnetic field component parallel to
the conductive direction, which would promote the motion 
along that direction, has unexpectedly little effect on the
3D butterfly.

So the puzzle is: why is the 3D butterfly so robust?  
The tight-binding model with distant hoppings cannot be 
reduced to the effective 2D Harper's equation as is done
here in going from eq.(\ref{Harp3D}) to eq.(\ref{HarpEff}), 
so we must take another approach.

We consider a 3D butterfly system ($t_x \gg t_y, t_z$)
in a magnetic field $(B_x,B_y,B_z)$.
We take the vector potential $\Vec{A} = (B_z y - B_y z, B_x z, 0)$,
so that $x$ is a cyclic coordinate and the wave function becomes
\begin{equation}
\psi_{lmn} = e^{i \nu_{x} l} G_{mn},
\end{equation}
where $(l,m,n)$ labels $(x,y,z)$.
Schr\"{o}dinger's equation becomes
\begin{eqnarray}
  &-&t_y(e^{i 2\pi \phi_x n} G_{m-1 n} \,+\, e^{-i 2\pi \phi_x n} G_{m+1 n})
  \nonumber \\
  &-&t_z(G_{m n-1} \,+\, G_{m n+1})   \nonumber \\
  &-&2t_x \cos[2\pi (\phi_z m - \phi_y n) +\nu_{x}] G_{mn} = E G_{mn},
\label{NewHarp}
\end{eqnarray}
where $\phi_{\mu}$ is the number of flux quanta penetrating 
the facet of the unit cell normal to the $\mu$ axis.

We first consider the case $B_x = 0$ ($\phi_x=0$).  
This is just the case we considered in the first 
part of this paper, but 
the equation cannot be reduced to a 2D Harper's equation due to 
the different gauge.  Nevertheless, we can derive an 
effective 2D Harper equation in the following manner.

In eq.(\ref{NewHarp}), the assumption $t_x \gg t_y,t_z$ means
the amplitude of the cosine potential, 
$t_x \cos[2\pi (\phi_z m - \phi_y n) +\nu_{x}]$, 
is much greater than the hopping terms along $y,z$ (first 
two terms in eq.(\ref{NewHarp})).  
So the states are basically bound to a 
`trough' of the cosine potential.  In this sense, the state
is quasi-one-dimensional. So let us derive an
effective Sch\"{o}dinger's equation within the trough.

If we switch off $t_z$ for the moment,
the set of equations for different $n$'s are decoupled to give
\begin{eqnarray}
  &-&t_y(G^{0}_{m-1 n} \,+\, G^{0}_{m+1 n}) \nonumber \\
  &-& 2t_x \cos[2\pi (\phi_z m - \phi_y n) +\nu_{x}] G^{0}_{mn} 
  = E^0_n G^{0}_{mn},
\label{NewHarp0}
\end{eqnarray}
This is an equation for one variable $m \parallel y$, and 
the eigenstates are bound states in a cosine potential.  If $t_z$ is
turned on, the states having different values of $n \parallel z$ 
are mixed to form a tight-binding band (Fig.\ref{trough}). 
The band width is proportional to $t_z$ times 
the overlap, $\alpha$, between neighboring bound states having 
a slightly different center positions. The equation
can now be reduced to
\begin{equation}
-\alpha t_z (J_{n-1} + J_{n+1}) + E^0_n J_n = E J_n,
\label{NewHarpEff}
\end{equation}
where $J_n$ is the amplitude of the bound states at $n$. 
The binding energy $E^0_n$ in eq.(\ref{NewHarp0}) 
depends on $n$, since the set of values of 
a steep cosine potential at discrete points have a 
discommensuration that is 
periodic in $n$ with a period $\phi_z / \phi_y$ 
(i.e., $n \rightarrow n + \phi_z / \phi_y$ in eq.(\ref{NewHarp0}) is
equivalent to $m \rightarrow m-1$).  
We can check by the numerical calculation that the 
$E^0_n$ can be approximated with a cosine function, 
whose amplitude is denoted by $t'$.  Thus, the equation
(\ref{NewHarpEff}) becomes
\begin{eqnarray}
&-&\alpha t_z (J_{n-1} + J_{n+1}) \nonumber\\
&-&2 t'\cos[2\pi (\phi_y / \phi_z) n + 
{\rm const.}] J_n = E J_n.
\label{NewHarpEff2}
\end{eqnarray}
This is the effective Harper
equation in this approach.  This equation has the same set of
eigenvalues as those of eq.(\ref{HarpEff}) as it should.  

If we assume $\alpha =1 (\phi_y \ll \phi_z)$, we can prove that
$t'$ in eqs.(\ref{NewHarpEff2}) and (\ref{HarpEff}) are 
identical.  In fact, eq. (\ref{NewHarpEff2}) has the same form as 
eq.(\ref{HarpEff}) except that $t'$
and $t_z$ are exchanged.  In the 2D Harper equation, 
exchanging the two
hoppings does not change the energy spectrum because of the gauge 
invariance, which concludes the proof.

The new effective equation can easily be extended to the
arbitrarily oriented magnetic fields.  If $B_x$ is turned on, the bound
states in a trough feel a magnetic field $B_x$ parallel to the 
$yz$ plane.  If the state were purely 
one-dimensional, it would not be affected by magnetic
fields.  So the problem is to what extent 
we can neglect the effect of $B_x$ on the quasi-one-dimensional 
state. 
We assume $\phi_y \ll \phi_z (\theta \approx 0)$ again. 
If one considers the state having an energy $E$ above the bottom,
the width of the wave function can be estimated as
$\sqrt{E/t_x}/\phi_z$, while the length scale over which 
the Peierls phase due to $B_x$ changes is $1/\phi_x$.
Thus we can say $B_x$ has a minor effect on the 3D butterfly when
\begin{equation}
\sqrt{\frac{E}{t_x}}<\frac{\phi_z}{\phi_x}.
\end{equation}
In the opposite limit, $\phi_y \gg \phi_z (\theta \approx 90^{\circ})$, 
we have only to replace $\phi_z$ with $\phi_y$ in the inequality.
In our example above, the condition, for the lowest Landau band, becomes 
$\phi_x < 3\phi_z (\theta \approx 0^{\circ}), 
 \phi_x < 3\phi_y (\theta \approx 90^{\circ})$, 
which is indeed satisfied in Fig.\ref{btflBx}.


\section{Relation to the semiclassical picture}
\label{SecSemiCl}

The semiclassical picture is often employed in 
examining the motion of electrons in magnetic field.  So 
this section addresses the question of whether or not the present approach 
in terms of Harper's equation can be interpreted in the semiclassical 
approach.  We shall conclude that the semiclassical approach
{\it cannot} describe the butterfly physics.

In the semiclassical approximation,
an electron in $k$-space moves along the
intersection of the equienergy surface with a plane 
perpendicular to $\Vec{B}$. If the orbit closes in $k$-space, 
we may apply Bohr-Sommerfeld's quantization condition,
\begin{equation}
  \left|\oint\Vec{k}_{\perp}\times{\rm d}\Vec{k}_{\perp}\right| =
    \frac{2\pi}{\ell_B^2}n,
\label{Quantum}
\end{equation}
which requires that the area 
enclosed by the orbit must be a multiple of $2\pi/\ell_B^2$, 
where $\ell_B=\sqrt{\hbar /eB}$ is the magnetic length.  

On the other hand, in Harper's equation, the binding in wells gives
the quantized states. Let us start with 
identifying the correspondence between
two quantizations, which look different at first sight.
In the following, we again assume the conditions
$\phi_y,\phi_z \ll 1, t_y\phi_z \gg t_z\phi_y$,
which are our starting point in deriving the effective Harper equation.

The energy dispersion, in the absence of magnetic fields, is
\begin{equation}
E(\Vec{k}) = -2t_x\cos k_x a -2t_y\cos k_y b -2t_z\cos k_z c,
\end{equation}
where $a,b,c$ are the lattice constants along $x,y,z$,
respectively. 
First we consider a semiclassical orbit around $\Gamma$ point, 
around which $E(\Vec{k})$ can be approximated, up to a constant, as
\begin{equation}
E(\Vec{k}) = t_x(k_x a)^2 + t_y(k_y b)^2 + t_z(k_z c)^2.
\end{equation}
An equienergy surface is then represented by an ellipsoid.
For $\Vec{B}$ tilted by an angle $\theta$ from
$z$ axis, the area enclosed by the orbit around $\Gamma$ 
having an energy $E$ is 
\begin{equation}
S = \frac{\pi E}{\cos\theta\sqrt{t_x a^2(t_y b^2 + t_z c^2 \tan^2\theta)}}.
\label{Area}
\end{equation}
Applying the quantization condition, we obtain the quantized energy,
\begin{equation}
E_n = 4\pi n\sqrt{t_x(t_y \phi_z^2 + t_z \phi_y^2)} .
\label{Elliptic}
\end{equation}

Around the $X$ point, $(k_x,k_y,k_z)=(0,0,\pi/c)$ on the edge
of the Brillouin zone, on the other hand, $E(\Vec{k})$ can be approximated as
\begin{equation}
E(\Vec{k}) = t_x(k_x a)^2 + t_y(k_y b)^2 - t_z(\pi - k_z c)^2.
\end{equation}
This time the equienergy surface is hyperbolic along $z$, while elliptic
along $x,y$. The area of an orbit encircling X is 
\begin{equation}
S = \frac{\pi E}{\cos\theta\sqrt{t_x a^2(t_y b^2 - t_z c^2 \tan^2\theta)}},
\end{equation}
and the quantized energy is 
\begin{equation}
E_n = 4\pi n\sqrt{t_x(t_y \phi_z^2 - t_z \phi_y^2)}.
\label{Hyperbolic}
\end{equation}


What corresponds to the quantized energy (\ref{Elliptic})
or (\ref{Hyperbolic}) in Harper's equation? 
The potential in Harper's equation consists of two periodic terms,
\begin{eqnarray*}
V^{(1)}(l) &=& -2t_y \cos(2\pi \phi_{z}l +\nu_{y}), \\
V^{(2)}(l) &=& -2t_z \cos(-2\pi \phi_{y}l +\nu_{z}).
\end{eqnarray*}
As discussed above, if the condition $t_y\phi_z \gg t_z\phi_y$
is satisfied, $V^{(1)}$ is responsible for 
potential peaks and dips while $V^{(2)}$
provides a slowly varying part.  The energy quanta are determined by the
curvature of the wells.  While the curvature depends mainly on $V^{(1)}$, 
$V^{(2)}$ also gives an additional contribution.  Let us consider the 
well around a minimum of $V^{(2)}$.  We can express the bottom of the
well in a harmonic form,
\begin{equation}
V^{(1)}(l) + V^{(2)}(l) = t_y (2\pi  \phi_z l)^2 + t_z (2\pi  \phi_y l)^2.
\end{equation}
The energy quantum is then given by
\begin{equation}
\hbar \omega_c = 4\pi \sqrt{t_x a^2(t_y \phi_z^2 + t_z \phi_y^2)},
\end{equation}
which is identical with eq.(\ref{Elliptic}).  
Around a maximum of $V^{(2)}$, the potential is written as
\begin{equation}
V^{(1)}(l) + V^{(2)}(l) = t_y (2\pi  \phi_z l)^2 - t_z (2\pi  \phi_y l)^2.
\end{equation}
Note that $V^{(1)}$ and $V^{(2)}$ have the opposite curvatures in this
case.  The energy quantum is 
\begin{equation}
\hbar \omega_c = 4\pi \sqrt{t_x a^2(t_y \phi_z^2 - t_z \phi_y^2)},
\end{equation}
which is identical with eq.(\ref{Hyperbolic}). 

These indicate that a semiclassical orbit in $k$-space translates
to a bound state in a well. In Harper's equation, however, a bound
state of one well is not an eigenstate but hops to adjacent 
wells, i.e.,  bound states in different wells are
mixed, which is exactly why each Landau level is broadened by 
the hopping $t'$.  
We have seen that this effect is essential in explaining 
the butterfly structure.  
Hence we can conclude that the butterfly is {\it outside} the 
semiclassical description.

\section{The Hall conductivity}
\label{SecHall}

Once the spectrum with a recursive set of energy gaps are obtained, 
a most important question is whether an integer quantum Hall 
effect exists on that spectrum.  
Montambaux and Kohmoto have first obtained 
the quantized Hall conductance for
three dimensional tight-binding electrons in magnetic fields
following the line of the results for the 2D case\cite{Mont}.
In this section, we derive,
following their arguments, the systematics in the Hall conductivity for 
the 3D butterfly spectrum.
Although the results are the same as in our earlier paper, 
we have greatly simplified the derivation.

Let us go back to Harper's equation (\ref{Harp3D}) for the simple
cubic lattice in rational magnetic fluxes
$(\phi_x,\phi_y,\phi_z)=(0,p_y/q_y,p_z/q_z)$.
When $\phi_z,\phi_y \ll 1$, 
we can adopt the effective-mass approach for 
the lattice structure along $x$, and the equation 
reduces to a differential form,
\begin{eqnarray}
E_x(-i\partial_x)F(x) 
\,-\, [2t_y \cos(G_z x  +k_y b) + \nonumber \\
2t_z \cos(-G_y x +k_z c)] F(x) = E F(x), 
\end{eqnarray}
where
\begin{equation}
G_z = 2 \pi \phi_z / a, \,  G_y = 2 \pi \phi_y / a, 
\end{equation}
and $E_x(k_x)$ is the dispersion along $x$ axis.
Now the equation is doubly periodic just as in 2D.
If we assume $\phi_z/\phi_y = p/q$ ($p,q$ :integers), 
we end up with a Diophantine equation for the $r$-th gap,
\begin{equation}
r = s q + t p,
\label{Dio3D}
\end{equation}
where $s, t$ are topologically invariant integers, constant in each gap.
Thus, the number of states per unit volume below the $r$th gap is
\begin{equation}
\frac{s G_z + t G_y}{2\pi/L_x}\, \frac{2\pi/b}{2\pi/L_y}\, 
\frac{2\pi/c}{2\pi/L_z} \frac{1}{V}
= \frac{s G_z + t G_y}{2\pi b c},
\label{NumOfStates}
\end{equation}
where $V = L_xL_yL_z$ is the total volume of the system.

According to the Widom-St\v{r}eda formula \cite{Wido,Stre}, 
the Hall conductivity when the Fermi level is in a gap is 
written as
\begin{equation}
\sigma_{ij} = -e \sum_{k} \epsilon_{ijk} \frac{\partial\rho}{\partial B_k},
\end{equation}
where $\rho$ is the density of electrons for a fixed chemical potential
and $\epsilon_{ijk}$ is the unit antisymmetric tensor.
Substituting eq.(\ref{NumOfStates}) for $\rho$, we finally obtain
\begin{equation}
(\sigma_{xy},\sigma_{zx}) = \frac{e^2}{h}\left( \frac{s}{c},\frac{t}{b}\right).
\end{equation} 

This is the explicit formula for the Hall conductivity derived from
Montambaux-Kohmoto's argument. 
We can then calculate the Hall conductivity for an arbitrary
gap in the 3D butterfly. The result for the 
Hall conductivity is shown in Fig.\ref{btflHall}(a) 
for the spectrum displayed in Fig.\ref{btflBx}(a).


If we compare this with the corresponding plot for 2D in
Fig.\ref{btflHall}(b), we can see that there exists a beautiful
correspondence between 2D and 3D cases: $\sigma_{zx}$ in 3D corresponds
to $\sigma_{\rm 2D}$ in 2D.  The mapping between 2D and 
3D Harper equations explains this in a
transparent manner.  Let us consider the Fermi level lying in the $r'$th
butterfly gap in the $N$th Landau level in the 3D butterfly.  Since one
Landau band is composed of $q$ butterfly bands, the total gap number $r$
from the bottom is $r = N q + r'$.  The Diophantine equation
(\ref{Dio3D}) then reads
\begin{equation}
r' = (s - N) q + t p.
\end{equation}
On the other hand, according to the argument by Thouless et al\cite{Thou}, 
the Hall conductivity in 2D for the flux $p_{\rm 2D}/q_{\rm 2D}$ is given by
\begin{equation}
\sigma_{\rm 2D} = -\frac{e^2}{h}t_{\rm 2D},
\end{equation}
where $t_{\rm 2D}$ is an integer given by the Diophantine equation
for the $r_{\rm 2D}$th gap,
\begin{equation}
 r _{\rm 2D}=s _{\rm 2D} q _{\rm 2D} + t _{\rm 2D} p _{\rm 2D}.
\label{Dio2D}
\end{equation}
Recalling the 2D-3D mapping, we can establish a correspondence
\begin{eqnarray}
p &\longleftrightarrow& p_{\rm 2D}, \nonumber \\
q &\longleftrightarrow& q_{\rm 2D}, \nonumber \\ 
r' &\longleftrightarrow& r_{\rm 2D}.
\end{eqnarray}
Namely, the correspondence between the topological invariants reads
\begin{eqnarray}
s - N &\longleftrightarrow& s_{\rm 2D}, \\
t   &\longleftrightarrow& t_{\rm 2D}. \label{CorresInteg}
\end{eqnarray}
This relation  explains the correspondence between
$\sigma_{zx}$ in 3D and $\sigma_{\rm 2D}$ in 2D as displayed 
in the Fig.\ref{btflHall}.

We can ask how the Hall conductivities vary when 
the magnetic field is tilted away from the $yz$ plane 
(with $B_x\neq 0$).  
As shown in the Sec.\ref{SecBx}, $B_x$ preserves the 
butterfly structure unless its magnitude is too large. This implies 
that the Hall conductivities, 
being topological invariants, are not affected by the tilting. 

\section{Possible experimental realizations}
\label{SecExp}

Let us comment on the possibility of observing the 3D butterfly 
in real materials.  In the 2D 
case, the magnetic flux $\phi$ required for the butterfly is
the order of unity, that is, $B \approx \frac{1}{a^2}\frac{h}{e}$.  If
one assumes an atomic lattice constant $a=2$\AA \, $B \sim O(10^5)$T, 
which is obviously too huge.  In the 3D case on the other
hand, Fig.\ref{phase} shows that there exists appropriate $(t_y,t_z)$
for the 3D butterfly no matter how small $\phi_y, \phi_z$ may be. So, for a
given lattice constant, the butterfly is more easily realized in 3D.

In practice, appropriate $(t_y,t_z)$ become small for small values of
the fluxes. The energy scale then shrinks when the magnetic field is
small, so that it will become harder to resolve the butterfly structure.
For typical quasi-1D organic conductors such as
(TMTSF)$_2$X\cite{ishiguro} we have $t_x:t_y:t_z\sim 1:0.1:0.01$ with
$a, b, c \sim 10 {\rm \AA}$, and we can estimate the required magnetic
field to be
\[
\phi_z \gsim 0.01,
\] 
which is {\it two} orders of magnitude smaller than the 2D 
counterpart, $\phi_z \sim 1$.  
$\phi_z = 0.01$ corresponds to $B = 40$T, which is well 
within the experimental feasibility. 
The energy scale (the required energy resolution) 
is $4(t'+t_z)\sim 10$ meV.  
In our earlier paper,\cite{Kosh} we have 
focused on the lowest Landau level in estimating the required 
field for the butterfly, which required a much larger flux ($\phi_z =
0.1$).  
So the important message here is that, 
if we employ higher Landau levels, the condition is drastically relaxed, 
since the higher levels have larger Harper-broadening.

The electron-electron interaction, ignored here, can be significant in
real materials.  Although it may first seem that the interaction would
smear out the butterfly, we can show that it can rather give rise to the
butterfly.  For an organic metal (TMTSF)$_2$X, it has been known that
the many-body interaction causes a spin density wave to become stable in
strong magnetic fields, which is called the field-induced spin density
wave (FISDW)\cite{Gorkov}. There, an energy gap opens across the SDW
nesting wave vector on the Fermi surface, and the hoppings $t_y$ and
$t_z$ give rise to series of gaps for electron and hole bands around the
SDW gap\cite{Poil,Maki}.  In (TMTSF)$_2$X where $t_x: t_y: t_z \sim 1:
0.1: 0.01$, $t_z$ happens to be too small to play a role, so that the
spectrum is 2D-like, i.e., has only Landau levels around the main gap.
We can then propose that, if there are quasi-1D materials that have $t_y
\approx t_z$ (rather than $t_y \gg t_z$), 
the spectrum is expected to have 3D butterfly gaps
according to the present theory. This mechanism is interesting, since
the butterfly formation, being SDW related, is around $E_F$ by
construction. The detail will be published elsewhere.

\section{Conclusions}
\label{SecConcl}

We have investigated the energy spectra and
quantum Hall effects in anisotropic three-dimensional Bloch electrons in
magnetic fields. We have (i) obtained a phase
diagram versus $(t_y, t_z)$ to identify the 
region in which the butterfly appears.  (ii) We
have extended our theory to the arbitrary orientation of $\Vec{B}$, and
proved that a change in the orientation has surprisingly little effect
on the 3D butterfly.  (iii) We have discussed the relation between our
method and the semiclassical approach, and found that while there is a
clear correspondence in Landau's quantization, the butterfly structure
is beyond the semiclassical picture.  (iv) The required magnetic
field for a representative organic metal (TMTSF compound) is estimated
to be modest ($\sim 40$ T) if we adopt higher Landau levels for the
butterfly.  (v) We have finally given a simpler way of deriving the
topological invariants that represent the quantum Hall numbers (i.e.,
two Hall conductivity in 3D, $\sigma_{xy}, \sigma_{zx}$, in units of
$e^2/h$).

M.K. would like to thank the JSPS Research Fellowships for Young
Scientists for financial support.

\begin{figure}
\caption{The geometry in a 3D lattice in a tilted magnetic field.}
\label{geom}
\end{figure}

\begin{figure}
\caption{The phase diagram that indicates the character of the
energy spectrum in anisotropic 3D systems $(t_y, t_z \ll t_x=1)$
against $t_y$ and $t_z$ 
for the given values of the magnetic fluxes $\phi_y, \phi_z$ 
as defined in Fig.\ref{geom}.  The 3D butterfly region is shaded.}  
\label{phase}
\end{figure}

\begin{figure}
\caption{Typical energy spectra plotted against ${\rm tan}\theta$ 
for various positions 
in the phase diagram (indicated on Fig.\ref{phase} 
as (a)-(e)). 
Here we have fixed $V^{(1)}$ 
($t_y = 0.5, \phi_z = 1/10$) to vary $t_z/\phi_y^2$.  
We can see that the butterfly region shifts upward 
as $t_z$ is increased, which confirms the 
above observation since the width of the Landau 
band is larger for higher Landau bands.
}
\label{btflMany}
\end{figure}

\begin{figure}
\caption{ Typical 1D-like energy spectrum for $t_x:t_y:t_z =
 1:0.01:0.01$ plotted against $\theta$ in magnetic fields
 $(\phi_y,\phi_z) = 1/5(\sin\theta,\cos\theta)$. The gaps are due to
 Bragg's reflection by $V^{(1)}$ and $V^{(2)}$.}
\label{1Dlike}
\end{figure}

\begin{figure}
\caption{
  The energy spectra of the 3D system $t_x:t_y:t_z
  = 1:0.1:0.1$ plotted against the tilting angle $\theta$ in a magnetic field
  $\sqrt{\phi_y^2+\phi_z^2} = 1/5$ 
  for various tilting orientations of $\Vec{B}$ (as 
  depicted in the inset). 
  }
\label{btflBx}
\end{figure}

\begin{figure}
\caption{The potential and wave functions along $y$ are 
schematically depicted for eq.(\ref{NewHarpEff}). The grayscale
represents the value of the potential. The eigenstates bound 
at different positions in the
trough are mixed by $t_z$.} 
\label{trough}
\end{figure}

\begin{figure}
\caption{
  Semiclassical electron orbits in
  $k$-space are schematically shown.  We have assumed $t_x > t_y > t_z$.}
\label{contour1}
\end{figure}

\begin{figure}
\caption{ (a) Hall conductivities $(\sigma_{xy},\sigma_{zx}) = -(e^2/h)
 (s, t)$ are plotted on a 3D butterfly, where we display the topological
 invariants $(s, t)$ for each gap.  (b) The corresponding plot for
 $\sigma _{\rm 2D} = -(e^2/h)t_{\rm 2D}$ on the 2D butterfly. The region 
 enclosed by a dashed line in (a) corresponds to the 2D butterfly, where
 $t$ in (a) corresponds to $t_{\rm 2D}$ in (b), while $s$ in (a) to
 $s_{\rm 2D}$ in Eq.(\ref{CorresInteg})} \label{btflHall}
\end{figure}

\end{multicols}
\end{document}